\documentclass[prl,twocolumn,showpacs,preprintnumbers,cite]{revtex4}
\usepackage{graphicx}% Include figure files
\usepackage{dcolumn}% Align table columns on decimal point
\usepackage{bm}% bold math

\usepackage{amssymb}
\usepackage{amsmath}

\begin{document}

\preprint{}

\title{Exotic Heavy-Fermion State in Filled Skutterudite SmOs$_4$Sb$_{12}$}

\author{Shotaro Sanada$^1$, Yuji Aoki$^1$, Hidekazu Aoki$^1$, Akihisa Tsuchiya$^1$, Daisuke Kikuchi$^1$, Hitoshi Sugawara$^2$, and Hideyuki Sato$^1$}
\affiliation{$^1$Department of Physics, Tokyo Metropolitan University, Hachioji, Tokyo 192-0397, Japan}% 
\affiliation{$^2$Faculty of Integrated Arts and Sciences, The University of Tokushima, Tokushima 770-8502, Japan}%

\date{\today}
%\date{\today}% It is always \today, today,%  but any date may be explicitly specified

\begin{abstract}
Specific heat and transport measurements have revealed an unconventional heavy-fermion (HF) state in SmOs$_4$Sb$_{12}$ single crystals.
The electronic specific-heat coefficient ($\gamma=0.82 $ J/K$^2$mol) and the coefficient ($A$) of the quadratic temperature dependence of electrical resistivity are largely enhanced, although the ratio $A\gamma^{-2}$ is reduced from the Kadowaki-Woods ratio of HF materials.
Both $\gamma$ and $A$ do not show any significant decrease in applied field in contrast with Ce-based HF compounds, suggesting an unconventional origin of the heavy quasiparticles.
A weak ferromagnetic ordering sets in below $\sim 3 $ K, probably originating in the itinerant quasiparticles.

\end{abstract}

%\pacs{74.70.Tx, 76.75.+i, 74.70.Dd, 74.25.Ha}% PACS, the Physics and Astronomy

% 74.70.Tx Heavy-fermion superconductors
% 76.75.+i Muon spin rotation and relaxation
% 74.70.Dd Ternary, quaternary and multinary compounds (including Chevrel phases, borocarbides, etc.)
% 74.25.Ha Magnetic properties SC

\maketitle
Filled skutterudite compounds, with a general formula RT$_4$X$_{12}$ (R=rare earth or U; T=Fe, Ru or Os; X=P, As or Sb) crystallizing in the unique bcc structure of the space group $Im \overline{3}$ (\#204)~\cite{Jeitschko1977} show a variety of interesting physical phenomena~\cite{Sales2003,AokiJPSJKondo}.
Among these, the Pr-based compounds PrFe$_4$P$_{12}$ and PrOs$_4$Sb$_{12}$ have attracted special attention because of their unusual heavy-fermion (HF) behavior.
PrFe$_4$P$_{12}$ exhibits HF behavior in magnetic fields~\cite{SatoPRB2000,AokiPFPPRB2002,SugawaraPFPPRB2002} in close proximity to an antiferroelectric-quadrupole (AFQ) ordering phase.
PrOs$_4$Sb$_{12}$ is the first-known Pr-based HF superconductor~\cite{BauerPOSPRB2002}, which shows unconventional superconducting behavior~\cite{IzawaPRL2003,AokiMSRPOSPRL2003} and an anomalous field-induced AFQ ordered phase above 4.5 T~\cite{AokiPOSJPSJ2002,TayamaPOSJPSJ2003,KohgiPOSJPSJ2003,RotunduPOSPRL2004}.
These findings of HF behavior in the Pr-based filled skutterudites are surprising since $4f$ electrons of Pr$^{3+}$ ions are generally considered to be quite stable (well localized) in intermetallic compounds.
The HF state in PrFe$_4$P$_{12}$ is quite unusual because no magnetic-moment screening is observed, in contrast with Ce- or Yb-based HF compounds where the magnetic Kondo effect plays a key role.
This fact, along with the AFQ phases appearing in these compounds, indicates that quadrupole interactions play an important role in the Pr-based filled skutterudites.

In this paper, we demonstrate that Sm ions also show strongly correlated behaviors in the filled skutterudites.
SmOs$_4$Sb$_{12}$ is a new HF material with the Sommerfeld coefficient $\gamma=0.82 $ J/K$^2$mol, which is the largest among the known Sm-based compounds.
Note that Sm-based HF behaviors have been found only in a few materials, e.g., SmPd$_3$ ($\gamma=0.28 $ J/K$^2$mol)~\cite{LiuJDP1988} and SmFe$_4$P$_{12}$ ($\gamma=0.37 $ J/K$^2$mol)~\cite{TakedaJPCM2003}.

Single crystals of SmOs$_4$Sb$_{12}$ were grown by the Sb-self-flux method using high-purity raw materials of 4N(99.99\% pure)-Sm, 4N-Os and 6N-Sb~\cite{comment1}.
No impurity phase has been detected in a powder X-ray diffraction pattern, except small amounts of Os and Sb metals of the order of $\sim 1 \%$ in volume.
The residual resistance ratio (RRR) is $\sim$13.
The lattice parameter was determined to be $a$=9.301 \AA, consistent with the reported value~\cite{BraunJLCM1980}.
Specific heat $C(H,T)$ for $H \parallel [100]$ was measured by a quasi-adiabatic heat pulse method described in ref.~\cite{AokiPOSJPSJ2002} using a dilution refrigerator equipped with an 8 T superconducting magnet. % \parallel, \langle100\rangle
The bulk magnetization $M(\mu_0H\le 7$\ T$,T\ge 2.0 $\ K$)$ was measured with a Quantum-Design superconducting quantum-interference device (SQUID) magnetometer.
The electrical resistivity was measured by the ordinary dc four-probe method, using a top-loading $^3$He cryostat with a 16 T superconducting magnet above 0.4~K.

\begin{figure} % Fig1
\includegraphics[scale=0.5]{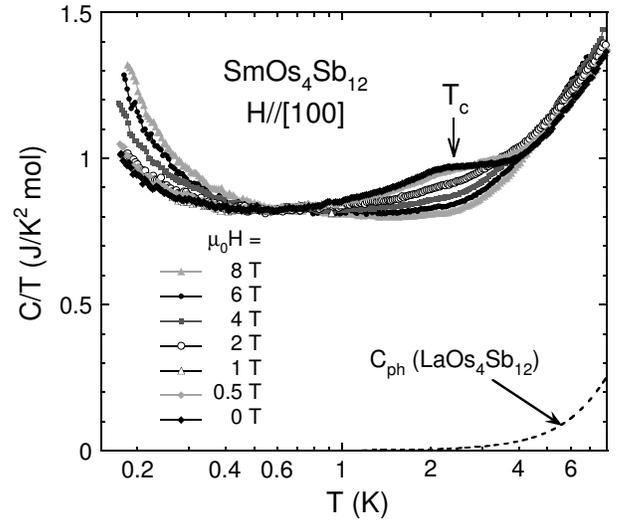}
\caption{Specific heat divided by temperature $C/T$ of SmOs$_4$Sb$_{12}$ measured in different magnetic fields. The broken curve represents the phonon part $C_{\rm ph}(T)$ determined from the $C(T)$ data of LaOs$_4$Sb$_{12}$~\cite{AokiPOSJPSJ2002}.}
\label{fig:CT}
\end{figure}

\begin{figure} % Fig2
\includegraphics[scale=0.5]{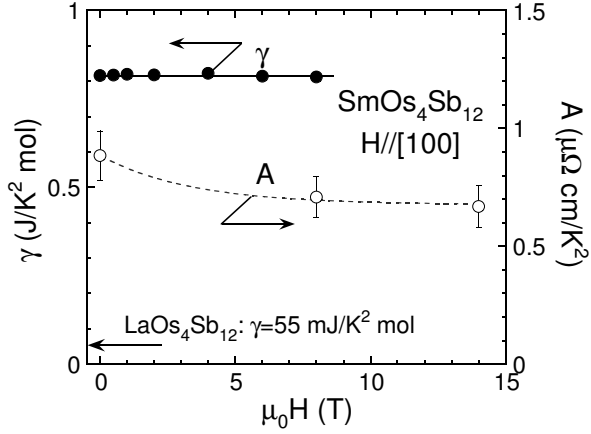}
\caption{Sommerfeld coefficient $\gamma$ and the coefficient $A$ of the quadratic $T$ dependence in the electrical resistivity $\rho$ of SmOs$_4$Sb$_{12}$ as a function of applied magnetic field $H \parallel [100]$. The $\gamma$ value of LaOs$_4$Sb$_{12}$ is indicated by the arrow.}
\label{fig:gamma}
\end{figure}

Figure~\ref{fig:CT} shows $C/T$ as a function of $\log T$ measured in applied field $H\parallel [100]$.
In this temperature range, $C$ consists of three components: the electronic ($C_{\rm e}$), nuclear Schottky ($C_{\rm n}$), and phonon ($C_{\rm ph}$) contributions.
An upturn below 0.5~K developing with increasing magnetic field is due to $C_{\rm n}$, which can be approximated as $A_{\rm n}/T^2$.
This term is dominated by Sb nuclear contribution; the zero-field value of $A_{\rm n}=1 \times 10^{-3}$ JK/mol is close to the expected value of $8.2 \times 10^{-4} $ JK/mol calculated based on the Sb-NQR measurement of PrOs$_4$Sb$_{12}$~\cite{KotegawaPRL2003}.
At 8~T, the experimental value of $A_{\rm n}=3.5 \times 10^{-3} $ JK/mol is close to the calculated Sb magnetic contribution of $3.6 \times 10^{-3} $ JK/mol.
Unfortunately, because of the small natural abundances and the weak onsite hyperfine coupling of $^{147}$Sm and $^{149}$Sm magnetic nuclei~\cite{BleaneyJAP1963}, one cannot obtain information on the 4$f$ magnetic moment of Sm ions.
Note that in PrFe$_4$P$_{12}$~\cite{AokiPFPPRB2002} and PrOs$_4$Sb$_{12}$~\cite{AokiPOSJPSJ2002}, $C_{\rm n}$ data have been utilized to clarify the nonmagnetic nature of the ordered state in the former compound and the magnetic behavior in the field-induced AFQ ordering phase in the latter compound.
In Fig.~\ref{fig:CT}, the increase in $C/T$ above $\sim 3 $ K can be explained by the phonon contribution ($C_{\rm ph}$) obtained from the LaOs$_4$Sb$_{12}$ data~\cite{AokiPOSJPSJ2002}. % , which is drawn in Fig.~\ref{fig:CT}.
We have obtained the electronic contribution $C_{\rm e}$ by subtracting $C_{\rm n}$ and $C_{\rm ph}$ from the experimental $C$ data.

The most significant finding shown in Fig.~\ref{fig:CT} is the almost temperature-independent $C_{\rm e}/T$.
This term dominates in $C/T$ in the temperature range between 0.3 and 3 K, where both $C_{\rm n}$ and $C_{\rm ph}$ are reduced.
The weak hump appearing at $T=2 \sim 3 $ K in zero field will be discussed below.
The Sommerfeld coefficient $\gamma$, obtained by fitting the data to $C(T) = A_{\rm n}/T^2+\gamma T$ below 1~K, is plotted in Fig.~\ref{fig:gamma} as a function of applied field.
The quite large $\gamma$ value of 0.82 J/K$^2$mol, 15 times larger than that for the no-$4f$ reference LaOs$_4$Sb$_{12}$ ($\gamma=0.055 $ J/K$^2$mol), suggests that the electron mass is highly enhanced by the many-body correlation effect.
Note that $\gamma$ is insensitive to the applied field, in contrast to the strong field dependence in the typical Ce-based HF compounds~\cite{BredlPRL1984,SatohSSC1985}.

\begin{figure} % Fig3
\includegraphics[scale=0.5]{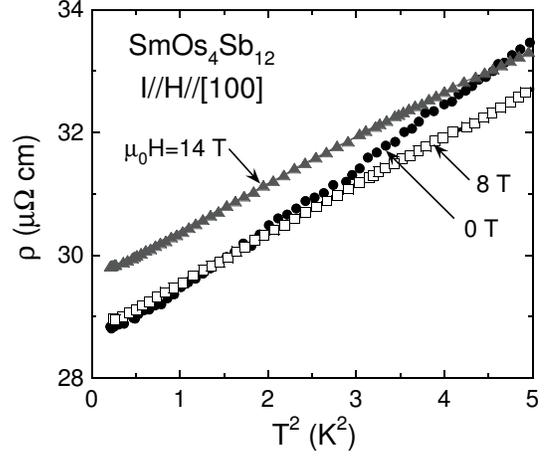}
\caption{Electrical resistivity $\rho$ vs $T^2$ in longitudinal magnetic field geometry.}
\label{fig:rho_average4}
\end{figure}

\begin{figure} % Fig4
\includegraphics[scale=0.5]{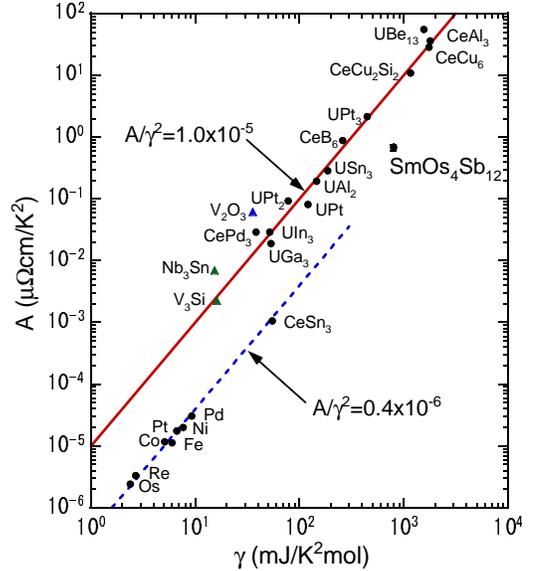}
\caption{Kadowaki-Woods plot ($A$ in $\rho=\rho_0+AT^2$ vs $\gamma$) with the data point for SmOs$_4$Sb$_{12}$. Data for other compounds are taken from the figures in refs.~\cite{KadowakiWoodsSSC1986} and~\cite{MiyakeSSC1989}.}
\label{fig:kw}
\end{figure}

The electrical resistivity $\rho$ measured in the longitudinal magnetic field geometry is shown in Fig.~\ref{fig:rho_average4}, where $\rho$ can be described well by $\rho=\rho_0 + AT^2$, indicating that electron-electron scatterings dominate at low temperatures.
The obtained coefficient $A$ is plotted in Fig.~\ref{fig:gamma}.
The slight enhancement of $A$ in zero field is probably due to the addition of ferromagnetic magnon scattering contribution since a weak ferromagnetic ordering sets in at $2 \sim 3$ K as discussed below.
This interpretation is also consistent with the negative magnetoresistance $\rho(8 $T$)<\rho(0 $T$)$. % for $T>1 $ K.
The positive magnetoresistance $\rho(14 $T$)-\rho(8 $T$)>0$ is ascribed to the ordinary contribution caused by the  electron's cyclotron motion.
If this contribution was subtracted from the data using Kohler's rule, the field dependence of $A$ seen in Fig.~\ref{fig:gamma} would become smaller.
Therefore, one can conclude that $A$ is as insensitive to the applied field as $\gamma$.

In the $A$ vs $\gamma$ diagram of Fig.~\ref{fig:kw}, the so-called Kadowaki-Woods plot, SmOs$_4$Sb$_{12}$ appears to be included in the group of HF materials~\cite{KadowakiWoodsSSC1986}, although the ratio $A \gamma^{-2}$ is reduced from the universal value of $1 \times 10^{-5}$ $\mu \Omega$cm(mol$\cdot$K/mJ)$^2$, which is discussed below.
This fact suggests that the largely enhanced values of $\gamma$ and $A$ are attributable to the Fermi liquid state, where heavy quasiparticles are formed. % undoubtedly

\begin{figure} % Fig5
\includegraphics[scale=0.5]{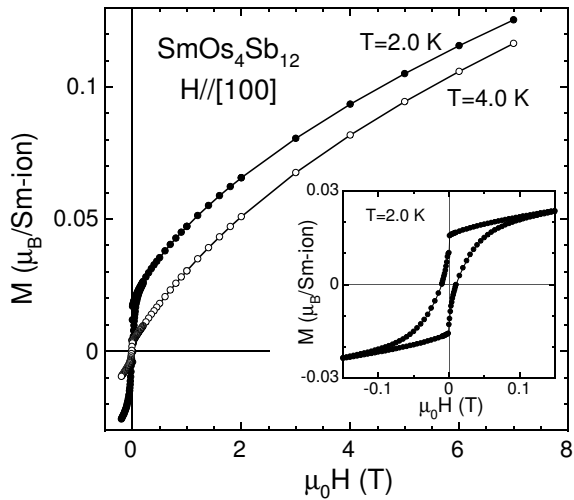}
\caption{Magnetization $M$ vs magnetic field $\mu_0 H$ isotherms at $T=$2.0 and 4.0 K.
Hysteretic behavior provides clear evidence of the ferromagnetic nature of the ground state.}
\label{fig:mh}
\end{figure}

\begin{figure} % Fig6
\includegraphics[scale=0.5]{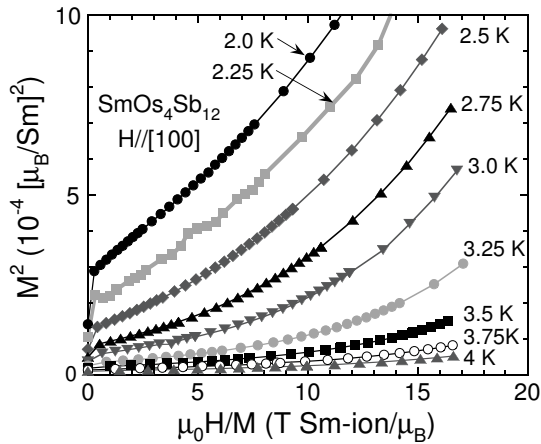}
\caption{Arrott plot $M^2$ vs $H/M$.}
\label{fig:ap}
\end{figure}

\begin{figure} % Fig7
\includegraphics[scale=0.5]{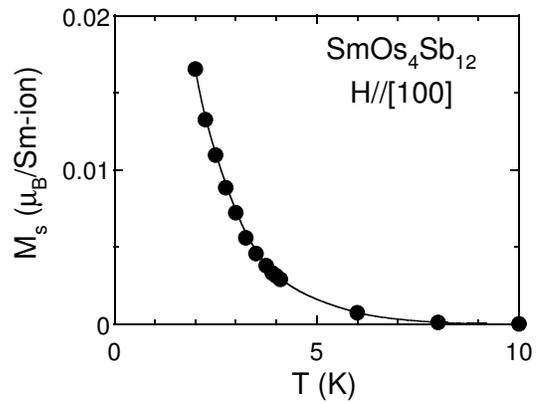}
\caption{Spontaneous magnetization determined by Arrott plot $M^2$ vs $H/M$ in Fig.~\ref{fig:ap}.}
\label{fig:ms}
\end{figure}

A magnetization $M$ vs $H$ isotherm measured at 2 K is shown in Fig.~\ref{fig:mh}.
A clear hysteretic behavior shows that a ferromagnetic ordering sets in at low temperatures~\cite{comment4}. % visible in this Figure 
In order to determine the ferromagnetic Curie temperature ($T_{\rm C}$) and the spontaneous magnetization ($M_{\rm s}$), we analyzed the measured magnetization curves using the Arrott plot~\cite{ArrottPR1957}.
The $M^2$ vs $H/M$ isotherms are plotted in Fig.~\ref{fig:ap}.
The temperature dependence of $M_{\rm s}$, provided by the $H/M \to 0$ intercept, is shown in Fig.~\ref{fig:ms}.
Compared with Fig.~\ref{fig:CT}, we conclude that the weak hump in $C(T)$ in the zero field appearing at $2 \sim 3 $ K ($=T_{\rm C}$) is due to the ferromagnetic transition.
With increasing field, the hump structure becomes broader and shifts to higher temperatures, which is the typical behavior expected for a ferromagnet.

In the well-localized $4f$-electron picture, the $J=5/2$ multiplet of Sm$^{3+}$ ions splits into a doublet $\Gamma_{5}$ ($\Gamma_{7}$) and a quartet $\Gamma_{67}$ ($\Gamma_{8}$) in the $T_h$ site symmetry (More familiar cubic $O$ labels are shown in the parentheses. See ref.~\cite{TakegaharaJPSJ2001} for details). 
The CEF level scheme in SmOs$_4$Sb$_{12}$ can be estimated using the experimentally determined CEF parameters for PrOs$_4$Sb$_{12}$ ($B^0_4=+0.0237 $ K and etc)~\cite{KohgiPOSJPSJ2003,GoremyINSPOScm} using the following extrapolation procedure.
The CEF parameters have the empirical form $B^m_l=\Theta_l \langle r^l \rangle A^m_l$, where Stevens' multiplicative factors $\Theta_l$ and the radial integrals over the unfilled $4f$ electrons $\langle r^l \rangle$ depend on the rare-earth ions (the values are given in refs.~\cite{Hutchings1964} and~\cite{FuldeAP1986}).
For a good approximation (in the point charge model), $A^m_l$ are independent of the particular rare-earth ions $R$ in the filled-skutterudite isostructure of $R$Os$_4$Sb$_{12}$~\cite{comment2}.
The scaled CEF parameter for SmOs$_4$Sb$_{12}$ is $B^0_4=-0.0546 $ K, predicting a quartet $\Gamma_{67}$ ground state and a doublet $\Gamma_{5}$ excited state with a small energy separation $\Delta_{\rm CEF}/k_{\rm B}=19 $ K. % in energy.
The strong $c$-$f$ hybridization effect in SmOs$_4$Sb$_{12}$ neglected above could modify the CEF level scheme to some extent from the present estimation.

\begin{figure} % Fig7
\includegraphics[scale=0.5]{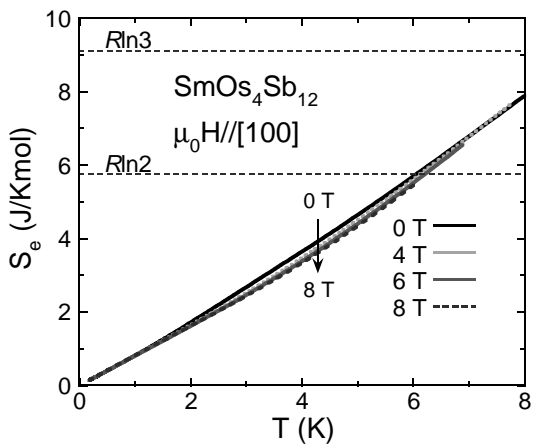}
\caption{Electronic part of entropy $S_{\rm e}(T)$ calculated integrating $C_{\rm e}(T)/T$ data.}
\label{fig:entropy}
\end{figure}

The electronic entropy $S_{\rm e}(T)$ calculated numerically integrating the $C_{\rm e}(T)/T$ data is shown in Fig.~\ref{fig:entropy}.
The insensitivity of $S_{\rm e}$ to the applied field is significant, as already seen in the $\gamma$ vs $H$ plot.
Without showing any tendency of saturation, $S_{\rm e}$ increases monotonically exceeding $R\ln2$ at 6~K.
This behavior agrees with the quartet $\Gamma_{67}$ ground-state model~\cite{comment3}.

In Ce-based HF compounds, like CeCu$_6$ and CeRu$_2$Si$_2$, $C_{\rm e}/T$ starts to decrease above around the Kondo temperature ($T_{\rm K}$).
In contrast, $C_{\rm e}/T$ of SmOs$_4$Sb$_{12}$ does not show any decrease with increasing $T$ (see Fig.~\ref{fig:CT}).
Therefore, the characteristic temperature of the HF state in SmOs$_4$Sb$_{12}$ (tentatively referred to as $T_{\rm K}$) is probably higher than 10 K, which is comparable with the estimated value of $\Delta_{\rm CEF}$.

The ratio $A \gamma^{-2}$ (in the unit of $\mu \Omega$cm(mol$\cdot$K/mJ)$^2$) of $1.1(2) \times 10^{-6}$ for SmOs$_4$Sb$_{12}$ is intermediate between $1 \times 10^{-5}$ for HF materials~\cite{KadowakiWoodsSSC1986} and $0.4 \times 10^{-6}$ for transition metals~\cite{RicePRL1968}.
While this may be attributed to the effect of moderate many-body correlations as discussed in ref.~\cite{MiyakeSSC1989}, the degeneracy due to the CEF effect could be another possible explanation.
A theoretical study in the orbitally degenerate periodic Anderson model suggests that $A \gamma^{-2}$ can be expressed roughly as $1 \times 10^{-5}/[N(N-1)/2]$, where $N$ is the orbital degeneracy, i.e., $A \gamma^{-2}$ becomes smaller for $f$-electrons with larger degeneracy~\cite{KontaniJPSJ2004}.
The reduced ratio in SmOs$_4$Sb$_{12}$ is close to the theoretically expected values of $1.7 \times 10^{-6} (N=4)$ or $0.7 \times 10^{-6} (N=6)$, suggesting the $f$-electron orbital degeneracy playing a role in the HF state.

The anomaly in $C_{\rm e}/T$ associated with the ferromagnetic transition is very weak.
The entropy released by the anomaly is only 2\% of $R \ln 2$.
Furthermore, the observed $M_{\rm s}$ is far below the expected values of 0.24 and 0.67 $\mu_{\rm B}/$Sm-ion for $\Gamma_5$ and $\Gamma_{67}$ CEF levels, respectively.
These features indicate that the ferromagnetic ordering originates in the itinerant heavy quasiparticles.

In summary, the specific heat and transport measurements have revealed the unconventional heavy-fermion state in SmOs$_4$Sb$_{12}$.
The Sommerfeld coefficient $\gamma=0.82 $ J/K$^2$mol and the coefficient $A$ of the $T^2$ dependence of electrical resistivity roughly satisfy the Kadowaki-Woods relation.
$\gamma$ and $A$ do not show any significant decrease in applied field, suggesting that the heavy quasiparticles have an unconventional origin.
The simple CEF analysis suggests a quartet ground state, which has both magnetic moments, probably responsible for the weak itinerant ferromagnetism, and electric quadrupole degrees of freedom~\cite{ShiinaJPSJ1997}.
Thus, it is possible that the quadrupole moments of Sm ions play a role in the anomalous HF behavior.
The present finding provides additional strong evidence that the filled skutterudite structure has unusual electronic states, where multi-$f$-electron ions can also show unprecedented strongly correlated electron behavior.

We thank H. Kontani, K. Miyake, O. Sakai and R. Shiina for stimulating discussions.
This work was supported by a Grant-in-Aid for Scientific Research Priority Area "Skutterudite" (No. 15072206) of MEXT, Japan.

%$^*$ Also at the Graduate University for Advanced Studies (SOKENDAI).
% , Kanagawa 240-0193, Japan.


\begin{thebibliography}{50}

\bibitem{Jeitschko1977} W. Jeitschko and D. Braun: Acta Crystallogr, Sect. B \textbf{33} (1977) 3401.

\bibitem{Sales2003} B. C. Sales: Handbook on the Physics and Chemistry of Rare Earths Vol. 33, ed. K.A. Gschneidner, Jr., J.-C.G. Bunzli and V. K. Pecharsky, (Elsevier Science B. V, 2003) p. 1-34.

\bibitem{AokiJPSJKondo} Y. Aoki, H. Sugawara, H. Sato and H. Harima: to be published in J. Phys. Soc. Jpn.

\bibitem{SatoPRB2000} H. Sato, Y. Abe, T. D. Matsuda, K. Abe, H. Sugawara and Y. Aok: Phys. Rev. B \textbf{62} (2000) 15125.

\bibitem{AokiPFPPRB2002} Y. Aoki, T. Namiki, T. D. Matsuda, K. Abe, H. Sugawara and H. Sato: Phys. Rev. B \textbf{65} (2002) 064446.

\bibitem{SugawaraPFPPRB2002} H. Sugawara, T.D. Matsuda, K. Abe, Y. Aoki, H. Sato, S. Nojiri, Y. Inada, R. Settai and Y. Onuki: Phys. Rev. B \textbf{66} (2002) 134411.

\bibitem{BauerPOSPRB2002} E. D. Bauer, N. A. Frederick, P.-C. Ho, V. S. Zapf and M. B. Maple: Phys. Rev. B, \textbf{65} (2002) 100506(R).

\bibitem{IzawaPRL2003} K.~Izawa, Y.~Nakajima, J.~Goryo, Y.~Matsuda, S.~Osaki, H.~Sugawara, H.~Sato, P.~Thalmeier and K.~Maki: Phys. Rev. Lett. \textbf{90} (2003) 117001.

\bibitem{AokiMSRPOSPRL2003} Y.~Aoki, A.~Tsuchiya, T.~Kanayama, S.R.~Saha, H.~Sugawara, H.~Sato, W.~Higemoto, A.~Koda, K.~Ohishi, K.~Nishiyama and R.~Kadono: Phys. Rev. Lett. \textbf{91} (2003) 067003.

\bibitem{AokiPOSJPSJ2002} Y. Aoki, T. Namiki, S. Ohsaki, S. R. Saha, H. Sugawara and H. Sato: J. Phys. Soc. Jpn. \textbf{71} (2002) 2098.

\bibitem{KohgiPOSJPSJ2003}  M. Kohgi, K. Iwasa, M. Nakajima, N. Metoki, S. Araki, N. Bernfoeft, J. M. Mignot, A. Gukasov, H. Sato, Y. Aoki and H. Sugawara: J. Phys. Soc. Jpn. \textbf{72} (2003) 1002.% 34)

\bibitem{TayamaPOSJPSJ2003} T. Tayama, T. Sakakibara, H. Sugawara, Y. Aoki and H. Sato: J. Phys. Soc. Jpn. \textbf{72} (2003) 1516.

\bibitem{RotunduPOSPRL2004} C. R. Rotundu, H. Tsujii, Y. Takano, B. Andraka, H. Sugawara, Y. Aoki and H. Sato: Phys. Rev. Lett. \textbf{92} (2004) 037203.

\bibitem{LiuJDP1988} B. Liu, M. Kasaya and T. Kasuya: J. de Physique Colloque \textbf{C8}, Supplement (1988) 369.

\bibitem{TakedaJPCM2003} N. Takeda and M. Ishikawa: J. Phys.: Condens. Matter \textbf{15} (2003) L229.

\bibitem{comment1} In single crystals of PrOs$_4$Sb$_{12}$ and LaOs$_4$Sb$_{12}$ grown by basically the same procedure as SmOs$_4$Sb$_{12}$, dHvA oscillations have been observed: H. Sugawara, S. Osaki, S. R. Saha, Y. Aoki, H. Sato, Y. Inada, H. Shishido, R. Settai, Y. Onuki, H. Harima and K. Oikawa: Phys. Rev. B \textbf{66} (2002) 220504(R).

\bibitem{BraunJLCM1980} D. J. Braun and W. Jeitschko: J. Less-Common Met. \textbf{72} (1980) 147.

\bibitem{KotegawaPRL2003} H.~Kotegawa, M.~Yogi, Y.~Imamura, Y.~Kawasaki, G.-q.~Zheng, Y.~Kitaoka, S.~Ohsaki, H.~Sugawara, Y.~Aoki and H.~Sato: Phys. Rev. Lett. \textbf{90} (2003) 027001.

\bibitem{BleaneyJAP1963} B.~Bleaney: J. Appl. Phys. {\bf 34} (1963) 1024.

\bibitem{BredlPRL1984} C. D. Bredl, S. Horn, F. Steglich, B. L\"uthi and R. M. Martin: Phys. Rev. Lett. \textbf{52} (1984) 1982.

\bibitem{SatohSSC1985} K. Satoh, T. Fujita, Y. Maeno, Y. \=Onuki, T. Komatsubara and T. Ohtsuka: Solid State Commun. \textbf{56} (1985) 327.

\bibitem{KadowakiWoodsSSC1986} K. Kadowaki and S. B. Woods: Solid State Commun. \textbf{58} (1986) 507.

\bibitem{MiyakeSSC1989} K. Miyake, T. Matsuura and C. M. Varma: Solid State Commun. \textbf{71} (1989) 1039.

\bibitem{comment4} The Hall resistivity shows a corresponding hysteretic behavior, which can be ascribed to the anomalous Hall effect caused by magnetic scatterings of conduction electrons by ferromagnetic moments. This fact suggests that the ferromagnetic anomaly is hardly attributable only to the inclusion of a small amount of a ferromagnetic impurity phase.

\bibitem{ArrottPR1957} A. Arrott: Phys. Rev. \textbf{108} (1957) 1394.

\bibitem{TakegaharaJPSJ2001}  K. Takegahara, H. Harima and A. Yanase: J. Phys. Soc. Jpn. \textbf{70} (2001) 1190.

\bibitem{GoremyINSPOScm} E. A. Goremychkin, R. Osborn, E. D. Bauer, M. B. Maple, N. A. Frederick, W. M. Yuhasz, F. M. Woodward and J. W. Lynn: Phys. Rev. Lett. \textbf{93} (2004) 157003.


\bibitem{Hutchings1964} M. T. Hutchings: Solid State Physics, eds. F. Seitz and D. Turnbull (Academic, New York, 1964), Vol. 16, p. 227.

\bibitem{FuldeAP1986} P. Fulde and M. Loewenhaupt: Adv. Phys. \textbf{34} (1986) 589.

\bibitem{comment2} The $A^m_l$ coefficients will also change, but it is expected that these effects are small since, in the point charge model, $A^m_4 \propto d^{-5}$ and $A^m_6 \propto d^{-7}$, where $d$ is the distance between rare-earth ions and their neighboring ions.

\bibitem{comment3} The possibility of a doublet $\Gamma_{5}$ ground state cannot be ruled out. If this is the case, a quartet excited state must be located closely in energy.
As demonstrated in the calculations for Ce-based HF systems [H.-U. Desgranges and J. W. Rasul: Phys. Rev. B, \textbf{36} (1987) 328. and Y. Shimizu and O. Sakai: J. Phys. Soc. Jpn. \textbf{65} (1996) 2632.], $c$-$f$ hybridization comparable with $\Delta_{\rm CEF}$ in energy can smear out the saturating behavior at $R\ln2$ in $S_{\rm e}(T)$, 


\bibitem{RicePRL1968} M. J. Rice: Phys. Rev. Lett. \textbf{20} (1968) 1439.

\bibitem{KontaniJPSJ2004} H. Kontani: J. Phys. Soc. Jpn. \textbf{73} (2004) 515.

\bibitem{ShiinaJPSJ1997} R. Shiina, H. Shiba and P. Thalmeier: J. Phys. Soc. Jpn. \textbf{66} (1997) 1741.

\end{thebibliography}
\end{document}